\documentclass[%
 reprint, 
 superscriptaddress,
 amsmath,amssymb,
 aps,
floatfix,
longbibliography
]{revtex4-1}

\usepackage{graphicx}
\usepackage{bm}

\usepackage{xcolor}
\usepackage[colorlinks=true,citecolor=blue,linkcolor=magenta]{hyperref}

\newcommand{\la}{\langle}
\newcommand{\ra}{\rangle}
\newcommand{\rar}{\rightarrow}

\newcommand{\ben}{\begin{eqnarray}}
\newcommand{\een}{\end{eqnarray}}

\newcommand{\be}{\begin{equation}}
\newcommand{\ee}{\end{equation}}

\begin{document}

\title{Nonadiabatic transitions in Landau-Zener grids:
integrability and semiclassical theory}
\author{Rajesh K. Malla}
\affiliation{Theoretical Division, and the Center for Nonlinear Studies, Los Alamos National Laboratory, Los Alamos, New Mexico 87545, USA}

\author{Vladimir Y. Chernyak}
\affiliation{Department of Chemistry, Wayne State University, 5101 Cass Ave, Detroit, Michigan 48202, USA}
\affiliation{Department of Mathematics, Wayne State University, 656 W. Kirby, Detroit, Michigan 48202, USA}
\author{Nikolai A. Sinitsyn}
\affiliation{Theoretical Division, Los Alamos National Laboratory, Los Alamos, New Mexico 87545, USA}

\date{\today}

\begin{abstract}
We demonstrate that the general model of a linearly time-dependent   crossing of two energy bands is integrable. Namely, the Hamiltonian of this model has a qaudratically time-dependent commuting operator. 
We apply this property to four-state Landau-Zener (LZ) models that have previously been used to describe the Landau-St\"uckelberg interferometry experiments with  an electron shuttling between two semiconductor quantum dots. 
The integrability  then leads to simple but nontrivial exact relations for the transition probabilities.
In addition, the integrability leads to a semiclassical theory that provides  analytical approximation for the transition probabilities in these models for all parameter values. The results predict a dynamic phase transition, and show that similarly-looking models  belong to different topological classes.
\end{abstract}

\maketitle

\section{Introduction}

The Landau-Zener (LZ) model describes an evolution for  amplitudes of two states with a time-dependent Hamiltonian
\be
H=\left( \begin{array}{cc}
b_1t & g\\
g^* & b_2t
\end{array}\right),
\label{lz}
\ee
where $b_{1,2}$ are called slopes of diabatic levels and $g$ is the inter-level coupling. The basis in which the off-diagonal elements of $H(t)$ are time-independent is called diabatic basis. 
The LZ formula provides an exact analytical expression for  the probability to remain in the same diabatic state
after the evolution during time $t\in (-\infty,+\infty)$:
\be
P_{LZ}=e^{-2\pi |g|^2/|b_1-b_2|}.
\label{lzf}
\ee
This formula plays a special role in the theory of nonadiabatic transitions because it can be used  as an approximation when the energy levels are mostly well separated. The adiabaticity is then broken only in disjoint  regions of time-energy, in which  the nonadiabatic dynamics is experienced only by pairs of states and the parameter time-dependence can be linearized.

For  nanoscale systems of modern interest, however, many states may experience the nonadiabatic transitions simultaneously, even when the linear  approximation of the parameter time-dependence near the nonadiabatic transitions is still applicable. The state evolution is then described by the nonstationary Schr\"odinger equation
\be
i\frac{d}{dt} |\psi\ra = H(t) |\psi\ra,
\label{shr-eq}
\ee
and the time-dependent Hamiltonian of a multistate Landau-Zener (MLZ) process has generally the form \cite{be}
\begin{equation}
 H(t)=Bt+A,
    \label{mlz0}
\end{equation}
where $A$ and $B$ are time-independent matrices, and $B$ is diagonal. Let $E$ be the diagonal part of $A$. The nonzero elements of $Bt+E$ are  called  diabatic energies and the corresponding eigenstates are called  diabatic states. As $t\rar \pm \infty$, the diabatic states coincide with the Hamiltonian eigenstates.
The goal of the MLZ theory is to find the amplitudes $S_{nm}$ and the transition probabilities, $P_{m \rar n}=\vert S_{nm}\vert^2$, from the diabatic states $m$ as $t \rightarrow -\infty$ to the states $n$ as $t \rar +\infty$.

  \begin{figure}[t!]
    \centering
    \includegraphics[width=0.45\textwidth]{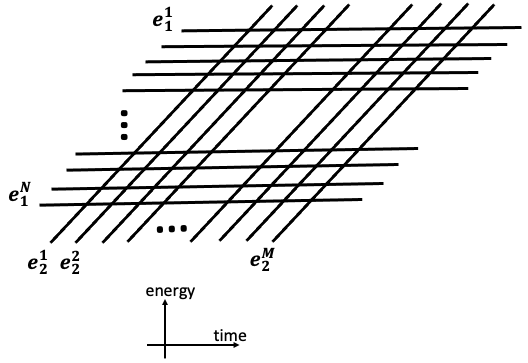}
    \caption{The linearly time-dependent diabatic levels in the model of two crossing bands are forming a pattern that we call LZ-grid. The first and the second bands have, respectively, $N$ and $M$ parallel  levels.  The  diabatic basis states of the same band do not interact with each other directly but any such a state can be coupled directly to arbitrary diabatic states of the other band. }
    \label{twobands-fig}%
\end{figure}
Among the MLZ models, there is a class of  Hamiltonians that has attracted special attention previously. It corresponds to the time-dependent crossing of two bands with parallel diabatic levels, as shown in Fig.~\ref{twobands-fig}. 
Let $N$ and $M$ be the integer numbers of the parallel levels in these bands.  Matrices $A$ and $B$ then have the dimensions $(N+M) \times (N+M)$, and 
\be
A = \left( 
\begin{array}{cccc}
E_1 & G \\
G^{\dagger} & E_2 
\end{array}
\right), \quad
B = \left( 
\begin{array}{cccc}
b_1{1}_N &0 \\
0&b_2{1}_M
\end{array}
\right),
\label{ham-gen1}
\ee
where $1_{N}$ and $1_M$ are the unit, respectively, N$\times$N and M$\times$M matrices.
The diagonal matrices 
$$
E_{1}\equiv{\rm diag}\{ e_{1}^1,e_{1}^2,\ldots, e_{1}^N\}, \,\,
E_{2}\equiv{\rm diag}\{ e_{2}^1,e_{2}^2,\ldots, e_{2}^M\}
$$
are responsible for the spacing between the parallel diabatic levels; $b_{1,2}$ are the slopes of the bands. The lower indices $1$ and $2$ in  $b_{1,2}^i,\, e_{1,2}^i$  refer, respectively, to the $N$-level and $M$-level bands.  $G$ is a $N\times M$ matrix that describes direct coupling between the two bands. All elements of $G$ can be nonzero and complex-valued.

The band crossing MLZ model (\ref{ham-gen1}) was discussed originally in relation to physics of the Rydberg atoms \cite{harmin1,harmin2}. Optical realization of this model was used to create an optical Galton board \cite{optics-lzgrid}. The early work sometimes referred to the level crossing pattern in the two-band model as to  {\it LZ-grid} of energy levels. 
Later, LZ-grids attracted attention in relation to the two-state systems  that are coupled to an environment, which splits the two    levels of the LZ model into the two bands  of many parallel levels  \cite{usuki,sinitsyn-bath,hanggie,garanin-grid,ashhab-lz}, 
and more recently LZ-grids emerged in the study of qubits coupled to optical modes, such as in circuit QED  systems \cite{osc-grid1,osc-grid2,osc-grid3,osc-grid4,recent-grid1,recent-grid2}.

Although certain facts about the LZ-grids  have been determined analytically \cite{usuki}, such systems remain generally unsolvable.  
Due to the complex oscillatory behavior of the transition probabilities as functions of the parameters, physics of LZ-grids remains poorly studied. Approximations have been developed but only for  limits of either very small \cite{ostrovsky-grid} or very large \cite{yurovsky-grid} separations of the parallel levels in the bands. Such limits are approached very slowly with decreasing/growing $e^i_{1,2}$, so they usually give too crude approximations to realistic choices of the parameters. There is also one fully solvable LZ-grid model \cite{quest-LZ} but it does not clarify many questions about the general problem. 

Our article has two goals. First, we add to the analytical understanding of the transition probabilities in all LZ-grids by applying the recent developments on the  time-dependent integrability \cite{commute,quadr-LZ}. Thus, in section~\ref{integrable-sec},  we show that all LZ-grids are formally integrable, in the sense that there is  an analytic expression for the time-quadratic-polynomial operator that commutes with the Hamiltonian, and satisfies an additional condition that is needed for the integrability of time-dependent Hamiltonians as defined in \cite{commute}.
However, unlike many known solvable cases \cite{laser-LZ,bcs,gamma-LZ,yuzbashyan-LZ}, LZ-grids  generally remain not fully solvable.  Instead, the integrability leads to a simple but nontrivial symmetry for the  transition probabilities.

Our second goal is to apply this symmetry to a specific four-state LZ-grid model that has attracted attention recently  due to experiments with Landau-St\"uckelberg  interferometry \cite{note} in coupled quantum dots \cite{petta-4lz1,petta-4lz2}. We will assume that the electrostatic energy of one of the dots is changing linearly with time and, in section \ref{fourstate-sec}, show that the integrability in such models leads to nontrivial exact relations between 
different transition probabilities. 

Finally, one of the applications of the integrability is the possibility to reduce the order of the differential equation for amplitudes of LZ-grid states. This property leads to  asymptotically exact expressions for  leading exponents that describe the transition probabilities in the nearly-adiabatic limit, as it has been recently demonstrated for the general three-state LZ model \cite{quadr-LZ}. 
In section~\ref{model1-sec} we apply this semiclassical approach to the  four-state quantum dot models in order to derive approximate expressions for the experimentally most relevant probability to remain in the same quantum dot after a linear sweep of the gate voltage.

\section{Integrability of the general band-crossing model}
  \label{integrable-sec}

 The integrability conditions
 for a time-dependent Hamiltonian $H(t)$ are defined as the possibility to find a parameter combination $\tau$, and an analytical form of a nontrivial operator $H'$ such that  \cite{commute}
\begin{eqnarray}
 \label{cond1}
 \frac{\partial H}{\partial \tau} - \frac{\partial H'}{\partial t} &=&0,\\
 \label{cond2}
 [H,H']&=&0.
\end{eqnarray} 
Such conditions are known in the theory of solitons \cite{faddeev-book},
and the existence of time-polynomial commuting operators in several MLZ systems was originally noticed in \cite{patra-LZ}.
Before we discuss the applications, let us first prove that the conditions (\ref{cond1}) and (\ref{cond2}) can be always satisfied for the LZ-grid Hamiltonian with $A$ and $B$ given by (\ref{ham-gen1}).

Let us  define a continuous family of operators 
\begin{equation}
 H(t,\tau)=B(\tau)t+A(\tau),
    \label{mlz-tau1}
\end{equation}
where $B(\tau)$ and $A(\tau)$ are obtained from the original $B$ and $A$ by setting
\be
B(\tau) \equiv B\tau, \quad E_1(\tau)\equiv \tau E_1,\quad G(\tau) \equiv G \sqrt{\tau},
\label{tau-intr1}
\ee
and keeping  $E_2$ intact.
Note that at $\tau=1$, $A(\tau)$ and $B(\tau)$ are the same as  original $A$ and $B$.

Then, the pair of operators, $H(t,\tau)$ and $H'(t,\tau)$, where
\be
H'(t,\tau) = \frac{\partial_{\tau}B(\tau)t^2}{2} +\partial_{\tau} A(\tau)t -\frac{1}{2(b_2-b_1)\tau^2} A^2(\tau),
\label{commute}
\ee
satisfy  (\ref{cond1}) and (\ref{cond2}).
Indeed, (\ref{cond1}) is trivial to verify, whereas (\ref{cond2}) leads to two independent conditions for the terms proportional to, separately, $t^2$ and $t$:
\begin{eqnarray}
\label{int-1}
&&\frac{1}{2}[\partial_{\tau}B(\tau),A(\tau)]+[\partial_{\tau}A(\tau),B(\tau)]=0,\\\label{int-2}
&&[\partial_{\tau}A(\tau),A(\tau)]-\frac{1}{2(b_2-b_1)\tau^2}[A^2(\tau),B(\tau)]=0,
\end{eqnarray}
which can  be verified by direct substitution of the $\tau$-dependent matrices (\ref{mlz-tau1}) and (\ref{tau-intr1}).

\begin{figure}[t!]
    \centering
    \includegraphics[width=0.45\textwidth]{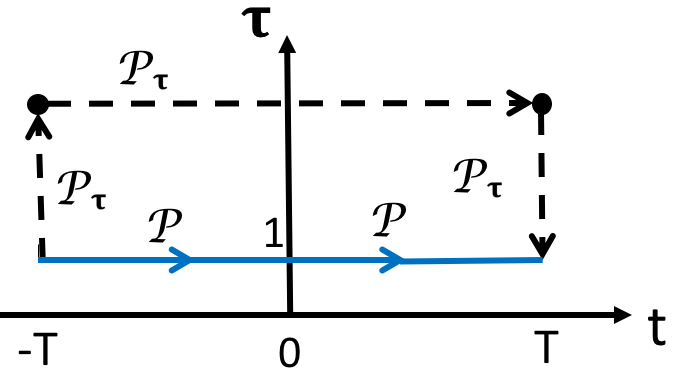}
    \caption{The true time-evolution path ${\cal P}$ (blue arrows) with $\tau=1$ and $t\in (-\infty,+\infty)$ can be deformed into the path ${\cal P}_{\tau}$, such that the horizontal part of ${\cal P}_{\tau}$ has $\tau = {\rm const} \ne 1$ (dashed black arrows). These deformations do not change the evolution matrix. Vertical legs of ${\cal P}_{\tau}$ have $t=\pm T $ with $T\rar \infty$, so they contribute only to the trivial adiabatic phases in the evolution matrix, and do not affect the transition probabilities. For the three-state LZ model, the path ${\cal P}_{\tau}$ can be chosen so that $b_1\rar \infty$ along the horizontal piece of this path. }
    \label{paths-fig1}%
\end{figure}

Due to the satisfied integrability conditions,  one can deform the integration path in the two-time space $(t,\tau)$ without changing the evolution amplitudes. Namely, let us define the evolution operator
 \be
 U=\hat{\cal T}_{\cal P}\exp \left( -i \int_{\cal P} H(t,\tau)\, dt+H'(t,\tau) \, d\tau \right),
 \label{path1}
 \ee
 where $\hat{\cal T}_{\cal P}$ is the path ordering operator along  ${\cal P}$ in the two-time space $(t,\tau)$.
 Equations~(\ref{cond1})-(\ref{cond2}) mean that the nonabelian gauge field with components ${\bf A}(t,\tau) = (H,H')$ has zero curvature, so the result of integration in (\ref{path1}) does not change after the deformations of ${\cal P}$  that keep only the initial and final points of ${\cal P}$ intact \cite{commute}, and avoid singularities of the $\tau$-dependent Hamiltonians, as in Fig.~\ref{paths-fig1}.

 Let the physical evolution correspond to the changes of $t$ from $-\infty$ to $+\infty$ at $\tau=1$. Then, ${\cal P}$ starts at the point $(t,\tau)=(-\infty,1)$.  We  fix, initially, $t$ and change $\tau$ from this point to another value, and only then perform $t$-evolution at fixed new $\tau$. After this, we   bring $\tau$ back to $\tau=1$ at $t=+\infty$ \cite{commute,faddeev-book}. 
 
 The $\tau$-evolution at fixed $t=-T \rar - \infty$ or $t=T\rar +\infty$ is strictly adiabatic due to the quadratic dependence of the diagonal elements of $H'$ on $t$. Therefore, the transition probabilities in the  LZ-grid models  that differ  only by  $\tau$ within the  family~(\ref{tau-intr1}), which is parametrized by $\tau$, are identical.

 This nontrivial invariance can now be used in combination with a trivial symmetry that is common for all MLZ models. Namely, by rescaling time in the Schr\"odinger equation (\ref{shr-eq}),
 \be
 t\rar t/\sqrt{\tau},
 \label{time-resc}
 \ee
 we cannot change the transition probabilities for the evolution in the interval $t\in (-\infty,\infty)$ at fixed $\tau$. On the other hand, this rescaling corresponds to the change of the parameters in the original model (\ref{ham-gen1}):
 \be
 b_{1,2}\rar b_{1,2}/\tau, \quad E_{1,2} \rar E_{1,2}/\sqrt{\tau}, \quad G \rar G/\sqrt{\tau}.
 \label{resc-2}
 \ee
 
Thus, the transition probabilities are independent of the variable transformations, simultaneously,  (\ref{tau-intr1}) and (\ref{resc-2}). Combining them, we find that the transition probabilities in  model (\ref{ham-gen1}) are invariant of a simple transformation of the diagonal matrices:
\be
E_1\rar E_1 \sqrt{\tau}, \quad E_2 \rar E_2 /\sqrt{\tau}.
\label{invar2}
\ee

\begin{figure}[t!]
   \centering
    \includegraphics[width=0.5\textwidth]{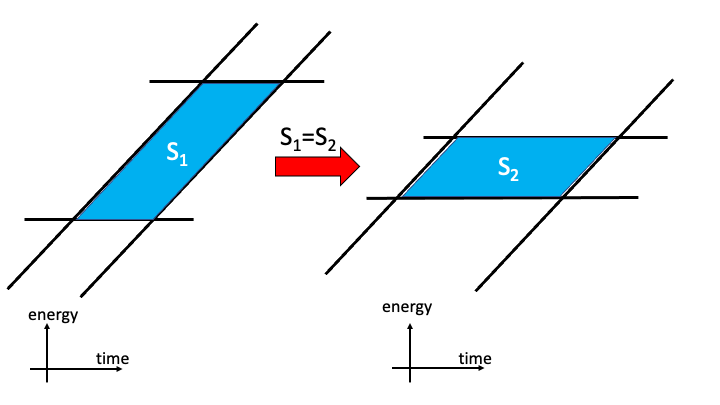}
    \caption{Varying the distances between parallel levels does not change the state-to-state transition probabilities if the area enclosed by the diabatic levels (filled by blue color) is conserved.}
    \label{areac}
\end{figure}

 This is the most general exact result of our article. The physical meaning of this result is illustrated in Fig.~\ref{areac}.  For a model with only two levels in each band in this figure, there are only two independent level splittings
 $$
 \Delta e_1\equiv e_1^1-e_1^2, \quad \Delta e_2\equiv e_2^1-e_2^2,
 $$
 so Eq.~(\ref{invar2}) means that
 the transition probabilities depend only on the combination $\Delta e_1 \Delta e_2$ but  not on the ratio $\Delta e_1/\Delta e_2$. It is easy to verify  that 
 $$
 S=\Delta e_1 \Delta e_2/(b_1-b_2)
 $$
 has the physical meaning of an area enclosed by the diabatic levels (the  diamond plaquette in Fig.~\ref{areac}). Hence, we can also formulate (\ref{invar2}) as an invariance of the transition probabilities of the transformations of the LZ-grid that preserve the areas enclosed by the diabatic levels, as well as the LZ parameters $|G_{ij}|^2/(b_1-b_2)$ if the slopes of the bands are also allowed to change.

Unfortunately, the invariance under the transformations (\ref{invar2}) is not sufficient to solve the whole model, i.e., to express the transition probabilities in terms of the known special functions of the model's parameters. Nevertheless,  we will show that this symmetry strongly simplifies the analysis, and even leads to certain further exact relations for the transition probabilities when a model has additional discrete symmetries.

\section{Quantum dot models}
\label{fourstate-sec}
In what follows, we will explore  application of the symmetry (\ref{invar2}) to two models, which have been studied to some extent  for various reasons previously.  We will refer to these models as to symmetric and antisymmetric. The symmetric model has the Hamiltonian
\be
H_{1}(t) = \left( 
\begin{array}{cccc}
e_1 & 0 & g &g \\
0 & -e_1 & g & g \\
g & g & bt+e_2 & 0 \\
g & g & 0 & bt-e_2
\end{array}
\right),
\label{ham-1}
\ee
and the antisymmetric one has the Hamiltonian
\be
H_{2}(t) = \left( 
\begin{array}{cccc}
e_1 & 0 & g &- \gamma \\
0 & -e_1 & \gamma & g \\
g & \gamma & bt+e_2 & 0 \\
-\gamma & g & 0 & bt-e_2
\end{array}
\right),
\label{ham-2}
\ee
where all parameters are real. The symmetric model has emerged previously in discussions of nonadiabatic behavior in MLZ systems in the large coupling  limit \cite{malla-LZ}. 

\begin{figure}[t!]
   \centering
    \includegraphics[width=0.4\textwidth]{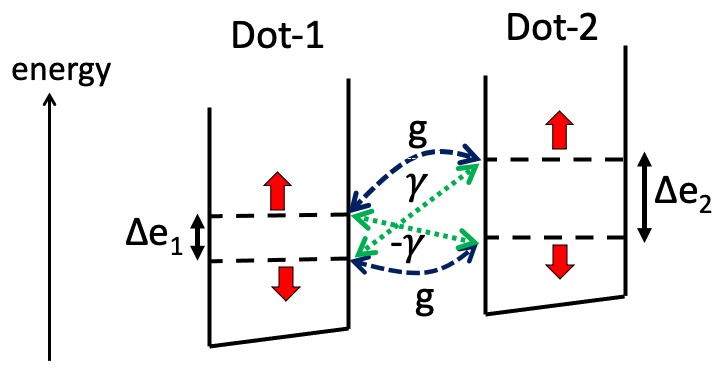}
    \caption{The antisymmetric model describes a single electron with spin shuttling between two quantum dots with the Hamiltonian (\ref{ham-2}). Dashed lines show pairs of energy levels in each dot. Up and down red arrows mark the orthogonal spin states. The splitting within each pair is the effect of an external static magnetic field. The relative energies of localized states in different dots are controlled by time-dependent electric gate voltage. The couplings $g$ and $\gamma$ describe, respectively, spin-conserving and spin-flipping electron tunnelings that are permitted by the time-reversal symmetry of the system in the absence of the magnetic field.  }
    \label{dots-fig}
\end{figure}

Physically, both models, (\ref{ham-1}) and (\ref{ham-2}), can describe a single electron that jumps between discrete levels of two quantum dots. The linear potential ramp in one of the dots is then induced by applying a time-dependent electric gate voltage difference between the dots, as in the experiments \cite{petta-4lz1,petta-4lz2}. Each dot has two discrete energy levels. 
Most naturally, this happens when electronic spin
can flip during the tunneling event due to the spin orbit coupling, as shown in Fig.~\ref{dots-fig}. 

In the antisymmetric case, with the Hamiltonian  $H_2$, the couplings $g$ and $\gamma$ describe then the spin preserving and the spin flipping tunneling between the two dots. The minus sign near $\gamma$ guarantees  the time-reversal invariance of the model. In fact, the Hamiltonian $H_2$ at $e_1=e_2=0$ is the most general, up to gauge transformations, Hamiltonian that one can create  for a time-reversed four-state system with spin \cite{four-LZ}, and the splittings $e_1,\,e_2\ne 0$ are induced by applying the external magnetic field. Generally, different quantum dots have different g-factors, which then means that $e_1\ne e_2$.

Both models, $H_1$ and $H_2$, have elementary discrete symmetries. 
For example, 
let
$$
\Theta=\left(\begin{array}{cccc}
0 &-1&0&0\\
-1 &0 &0&0 \\
0&0&0&1\\
0&0&1&0
\end{array}\right).
$$
The Hamiltonian $H_1$ has an elementary symmetry
$$
H_1(t)=-\Theta H_1(-t) \Theta,
$$
which leads to the symmetry of the evolution operator $U(T|-T)={\cal T}\exp(-i\int H_1(t)\,dt)$:
\be
U=\Theta U^{\dagger} \Theta,
\label{sym1}
\ee
from which follows that the amplitudes $U_{12}$ and $U_{43}$ in the symmetric model are purely real. There are also 
 relations between different scattering amplitudes, such as, $U_{13}=-U_{42}^*$, $U_{14}=-U_{32}^*$, which lead to the  relations between the transition probabilities: $P_{3\rar 1}=P_{2\rar 4}$, $P_{4 \rar 1}=P_{2\rar 3}$, $P_{3\rar 2}=P_{1\rar 4}$, and $P_{4\rar 2}=P_{1\rar 3}$. For a reader interested in examples of discrete symmetry effects in other MLZ models, we refer to Refs.~\cite{four-LZ,li-dynamic}.

Similarly, for the antisymmetric model, let
$$
\Theta_A=\left(\begin{array}{cccc}
0 &-i&0&0\\
i &0 &0&0 \\
0&0&0&i\\
0&0&-i&0
\end{array}\right),
$$
then
$$
H_2(t)=-\Theta_A H_2(-t) \Theta_A,
$$
from which follows that $U_{12}$ and $U_{43}$ are purely imaginary in this model, and there is the same set of relations between the transition probabilities as for the symmetric model.

\begin{figure}[t!]
    \centering
    \includegraphics[width=0.45\textwidth]{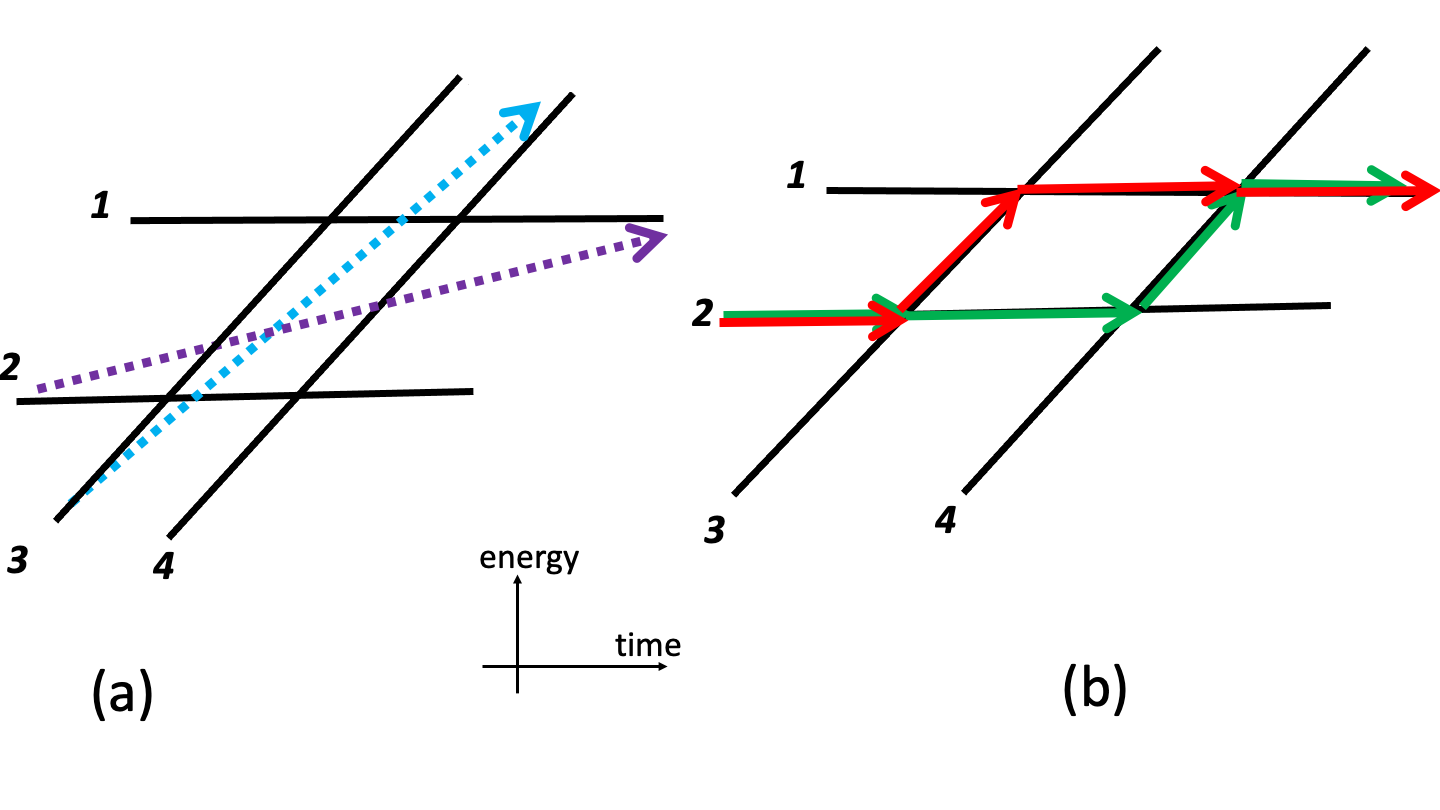}
    \caption{Diabatic levels of the quantum dot models. (a) Dashed colored arrows  show the transitions from level 2 to level 1 (violet) and from level 3 to level 4 (blue) with equal transition probabilities in both the symmetric and the antisymmetric models. Due to the difference, $e_1\ne e_2$, there is no obvious geometric symmetry between such transitions. (b) Two semiclassical paths in the diabatic level diagram that contribute to the transition probability from level 2 to level 1 (red and green arrows). Due to the quantum interference, this probability is expected to show oscillatory dependence on the area enclosed by the diabatic levels, $S=e_1e_2/b$.
      }
\label{p12-fig}      
\end{figure}

Apart from this, the  transition probability independence of  the area conserving transformations (Fig.~\ref{areac}), leads to  less intuitive  constraints. Namely,  since different quantum dots generally have different level splittings, i.e., 
$$
e_1\ne e_2
$$
there is generally an asymmetry of dynamics in respect to the initially chosen quantum dot. For example, the trivial symmetry (\ref{sym1}) does not predict any relation between the probabilities $P_{2\rar 1}$ and $P_{3\rar 4}$, as we illustrate in Fig.~\ref{p12-fig}(a).

However, the invariance of the transition probabilities of the transformations (\ref{invar2}) allows us to tune $e_1=e_2$ in both models without affecting the transition probabilities. Let us also add the gauge transformation $H_{1,2} \rar H_{1,2}-b 1_4/2$, which does not change the probabilities either. 
At such values of the parameters, the Hamiltonians (\ref{ham-1}) and (\ref{ham-2}) have an additional  discrete symmetry. Namely, let
$$
\theta=\left(\begin{array}{cccc}
0 &0&0&-1\\
0 &0 &-1&0 \\
0&1&0&0\\
1&0&0&0
\end{array}\right).
$$
Then, $H_{1,2}'=H_{1,2}^{e_1=e_2}-b 1_4/2$ satisfy
$$
H'_{1,2}(t)=\theta H'_{1,2}(t) \theta,
$$
from which we obtain additional relations on the transition amplitudes, such as $U_{12}=U_{43}^*$ and consequently $P_{2\rar 1}=P_{3\rar 4}$, e.t.c.. Let us now summarize all the relations among the transition probabilities that follow from simultaneously the integrability and the discrete symmetries of the models. Although the latter symmetries are a bit different for $H_1$ and $H_2$, they lead to the same relations between the transition probabilities:
\begin{eqnarray}
\label{rel1}
P_{1\rar 3} &=& P_{4\rar 2}, \\
\label{rel0}
P_{3\rar 4}&=&P_{2\rar 1},\\
\label{rel2}
P_{3\rar 1} &=& P_{2\rar 4},\\
\label{rel3}
P_{1\rar 4} &=&P_{2\rar 3}=P_{3\rar 2}=P_{4\rar 1}.
\end{eqnarray}
\begin{figure}[t!]
   \centering
    \includegraphics[width=0.45\textwidth]{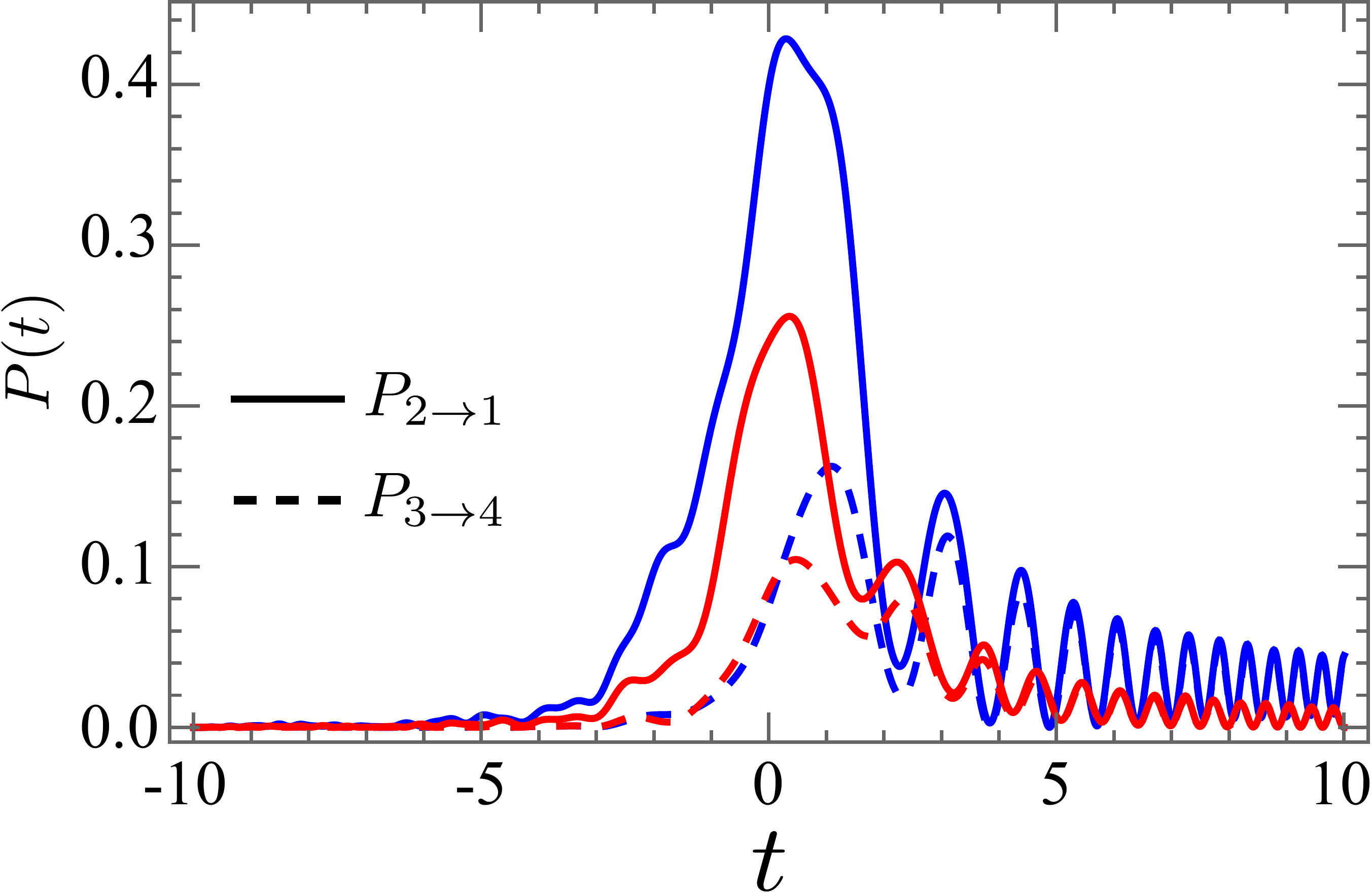}
    \caption{Numerically found time-dependence of transition probabilities $P_{2\rightarrow 1}$ (solid curves) and $P_{3\rightarrow 4}$ (dashed curves) in the antisymmetric model  for $\{e_1,e_2\}$=$\{1,3\}$ (blue),  $\{1,2\}$ (red). The remaining parameters: $b=2$, $g=2$, and $\gamma=1$. As both $P_{2\rar 1}$ and $P_{3\rar4}$ saturate at the same values at $t=\pm \infty$, their intermediate time dynamics are different. }
    \label{timeplot-fig}
\end{figure}

While for $e_1=e_2$, the relations (\ref{rel1})-(\ref{rel3}) are consequences of trivial discrete symmetries, for $e_1\ne e_2$ they are generally the results of the integrability of the LZ-grid model. We also note that our analysis, and hence relations~(\ref{rel1})-(\ref{rel3}) apply only to evolution from $t=-\infty$ to $t=+\infty$, whereas they do not apply   to the probabilities at intermediate times, except for $ e_1= e_2$, as we illustrate in Fig.~\ref{timeplot-fig}.
Hence, Eqs.~(\ref{rel1})-(\ref{rel3}) are our first nontrivial prediction for experimental verification. 

The  integrability of time-dependent Hamiltonians is a type of quantum symmetries that have not been studied experimentally previously. The experiments on Landau-St\"uckelberg interferometry provide an opportunity to detect the presence of such unusual quantum symmetries by measuring the state-to-state transition probabilities in already available solid state and atomic systems. The deviations from the exact predictions would mean the presence of the terms beyond the standard Hamiltonian $(\ref{ham-2})$, which may emerge either due to nonlinear time-dependence of the gate voltage or  effects of the magnetic field on the tunneling amplitudes.

In addition to the integrability conditions, there are six elements of the transition probability matrix that are known exactly and explicitly due to the no-go rule and the Brundobler-Elser formula \cite{be,usuki,nogo-LZ}. For the antisymmetric model they are
\begin{eqnarray}
\label{nogo1}
P_{1\rar 2} &=& P_{4\rar 3} = 0,\\
\label{be-eq}
P_{1\rar 1} &=&P_{2\rar 2} =P_{3\rar3} = P_{4\rar 4} =e^{-2\pi(g^2+\gamma^2)/b},
\end{eqnarray}
and for the symmetric model we should replace $\gamma \rar g$.

In addition to relations (\ref{rel1})-(\ref{rel3}), (\ref{nogo1}), and (\ref{be-eq}), 
we can only add the unitarity of the evolution constraints:
$$
\sum_{n=1}^4 P_{m\rar n}=\sum_{n=1}^4 P_{n\rar m}=1, \, \forall m\in \{1,2,3,4\}.
$$
It turns out that many of the latter relations are not independent after we include the already mentioned relations. Thus, even having so many constraints, the matrix of the transition probabilities has three unknown independent parameters that have to be calculated separately. 

Fortunately, the integrability leads to another simplification of the model's analysis. Namely, it was shown in \cite{quadr-LZ} that the integrability enables a semiclassical approach for estimation of the transition probabilities in the nearly-adiabatic limit. It was also noted in \cite{quadr-LZ} that the analytical formulas that are obtained by this approach often provide a reasonable approximation for the numerical solutions at arbitrary values of the parameters. 

Hence, in the following sections we apply this semiclassical approach to our four-state models, $H_1$ and $H_2$. For simplicity, here we will restrict ourselves only to the transition probability $P_{2\rar1}=P_{3\rar 4}$. This transition probability is the most physically interesting, first, because it is the only one that is needed to estimate the probability to remain in the same quantum dot after the time-linear sweep of the gate voltage. Indeed, all the other needed for this probabilities are given by the exact expressions (\ref{nogo1}) and (\ref{be-eq}). Moreover, a simple analysis shows that the probability $P_{2\rar1}$ should generally dominate over $P_{1\rar 1}=P_{2\rar 2}$ in the adiabatic limit because the latter become nonzero  after two, rather than one in the case of $P_{2\rar 1}$, nonadiabatic overgap transitions. 

Second, the transition from level 2 to level 1 is nontrivial even in the limit of large separation of all level crossings because it is then influenced by quantum interference of different evolution trajectories, as we show in Fig.~\ref{p12-fig}(b). 
The dependence of the probabilities of such transitions on the relative parameter values is understood poorly. Although numerical simulations of few-state systems are easy, they show complex oscillatory behavior on the parameters even in simplest and perturbative regimes \cite{kiselev-LZ}, 
so it is hard to see a general pattern for the role of different parameters. Hence, by developing nonperturbative analytical description of $P_{3\rar 4}$ in models  (\ref{ham-1}) and (\ref{ham-2})
we will obtain a useful insight into the interference effects in the nonadiabatic regime.


\section{Semiclassical solution for $P_{2\rar 1}=P_{3\rar 4}$ in quantum dot models}
\label{model1-sec}
In order to find the transition probability $P_{3\rightarrow 4}$  for models  (\ref{ham-1}) and (\ref{ham-2}) in the adiabatic limit,  we take advantage of the  $\tau$-independence of the transition probability and make the slopes of the tilted levels, 3 and 4, infinite by setting $\tau\rightarrow \infty$, as it was done to study the three-state MLZ model in \cite{quadr-LZ}. The tilted levels cross both the levels 1 and 2 then at time moments:
$$
t_{\pm}= \pm e_2/b,
$$
where ``$-$" is for the crossing of level 3 and ``$+$" is for the crossing of level 4.

 Let us assume that  level 3 is initially populated and we want to find the probability to end up on level 4.  Following \cite{quadr-LZ}, we can employ the fact that  for high-slope crossing the characteristic time of the nonadiabatic interactions is vanishing as $\delta t \sim 1/\sqrt{\tau}$, whereas the nonadiabatic transitions between the diabatic states 3 and 4 take time that is independent of $\tau$. Hence, we can separately treat the interactions of levels 3 and 4 with level 1, then with each other, and then with level 2. Specific details, however, depend on the models, which we will consider separately.

\subsection{Symmetric model}
First, we consider the Hamiltonian $H_1(t,\tau)$ in the limit $\tau \rar \infty$, and where $\tau$-dependence is according to (\ref{tau-intr1}).
We introduce the symmetric, $|+\rangle$, and the anti-symmetric, $|-\rangle$, combinations of the diabatic states 1 and 2: 
\be
\vert \pm \ra=\frac{1}{\sqrt{2}} \left( \vert 1 \ra \pm \vert 2 \ra \right).
\label{pm-states}
\ee
As $\tau \rar \infty$,  both levels 3 and 4 couple only to the $|+\ra$ state and
the nonadiabatic transitions to and from $|+\ra$ happen in the direct vicinity of time moments  $t_{-}$ and $t_{+}$. The corresponding coupling is $g\sqrt{2\tau}$, and the slope difference between the diabatic levels of $|+\ra$ and either $|3\ra$ or $|4\ra$ is $b\tau$. The probabilities of fast transitions from  $|3\ra$ to $|+\rangle$ near $t_{-}$ and from $|+\rangle$ to $ |4\ra$ near $t_+$ are given by the LZ formula for two state transitions:
$$
P_{3\rar +}=P_{+\rar 4} = 1-e^{-4\pi g^2/b}.
$$

In addition, during the time interval $t\in(t_{-},t_{+})$ the sates $|+\ra$ and $|-\ra$ interact with each other, in particular,  via the virtual transitions trough $|1\ra$ and $|2\ra$, as explained in \cite{quadr-LZ}. 
Hence, the total transition probability from level 3 to level 4 can be expressed via the product of the probabilities
\be
P_{3\rar 4}=P_{3\rar +}P^{+ +}P_{+\rar 4}=\left( 1-e^{-4\pi g^2/b}\right)^2 P^{+ +},
\label{interm-form1}
\ee
where  $P^{+ +}$ is the probability to remain in  $|+\ra$ after the dynamics with a $2\times 2$ effective Hamiltonian that acts in the subspace of $|\pm\rangle$ during the time interval $t\in (t_{-},t_{+})$. 

Away from the points $t=t_{\pm}$, as $\tau \rar \infty$, the effect of virtual transitions to and from levels 1 and 2 can be calculated perturbatively. Up to the zeroth order in $\tau$, such transitions lead to an effective   Hamiltonian in the subspace of states $|\pm \ra$:
\be
h_{v} = \frac{-2g^2 }{b}\left(\frac{1}{t-t_{-}} +\frac{1}{t-t_{+}}  \right)\vert +\ra \la +\vert,
\label{decay-term}
\ee
where the index $v$ refers to the interactions appearing due to the  virtual transitions in the second order of perturbation series over $1/\tau^{1/2}$. The higher order corrections due to such transitions in the limit $\tau \rar \infty$ vanish.

The other contribution to the effective two-state Hamiltonian arrives from the splitting  $ e_1$  of levels 1 and 2, which mixes states $\vert \pm \ra$:
\be
h_{e} = e_1 \left( \vert - \ra\la +\vert + \vert + \ra \la - \vert \right) .
\label{coupl-term}
\ee
$P^{++}$ can be determined by solving the effective Schr{\" o}dinger equation with the Hamiltonian $H_{\rm eff}=h_v+h_e$. After rescaling  time $t\rar t e_2/b$, this equation in the basis of states $|\pm \ra$ is given by
\be
i\frac{d}{dt} |\psi \ra =\frac{1}{b}  
H_{\rm eff}^s (t) |\psi \ra, \quad t\in (-1,1),
\label{eq-eff1}
\ee
where 
\be
H_{\rm eff}^s (t) = \left( 
\begin{array}{cc}
\frac{4g^2t}{t^2-1} & e_1e_2 \\
e_1e_2 & 0
\end{array}
\right).
\label{heff-sym}
\ee
In order to find $ P^{++}$, we should find the amplitude of evolution from $|+\ra$ as $t\rar -1$ into $|+\ra $ as $t\rar +1$, and take its absolute value squared. Note that the parameter $b$ plays the role in (\ref{eq-eff1}) of an effective Planck constant. 

Thus, the integrability reduces the problem to   a two-state system with time-dependent 2$\times$2 Hamiltonian $H_{\rm eff}^s$. 
It is possible to reformulate equation~(\ref{eq-eff1}) by changing to the variable $x \in (-\infty, +\infty)$, where $t=\tanh(x)$, and $dt/(t^2-1)=dx$. This  makes  it standard for application of the, so-called, Dykhne formula for the overgap transition probability between two states \cite{approx-LZ5,approx-LZ0}  in the adiabatic limit, in which   overgap transitions are suppressed  exponentially. This formula provides  the corresponding slowest decaying exponent and its leading order prefactor. 
We will use  this formula without switching from $t$ to $x$ because the direct application of  the Dykhne formula to the evolution during $t\in(-1,1)$ produces the same final result, and can be equally justified for the evolution (\ref{eq-eff1}).


The difference of the eigenvalues of $H_{\rm eff}^s$  is found analytically:
\be
\Delta E(t)  =2\frac{\sqrt{(e_1e_2)^2(t^2-1)^2+4g^4 t^2}}{(1-t^2)}.
\label{en-dif1}
\ee
By equating this difference to zero we find the branching points:
\be
t_{1,2}= \left(\frac{-2+r^2 \pm 2\sqrt{1-r^2}}{^2}\right)^{1/2}, \quad r\equiv \frac{(e_1e_2)}{g^2}.
\label{t12}
\ee
where the square root convention in $(\ldots)^{1/2}$ is chosen so that ${\rm Im}(t_{1,2}) >0$, and $t_1$ corresponds to the ``$+$" sign in (\ref{t12}). Depending on whether the ratio $r$ is bigger or smaller than $1$, we found two different types of the behavior.

{\bf Phase I}:  $r<1$, i.e., $e_1 e_2< g^2$. In this case, $\sqrt{1-r^2}$ is real, so
the branching points $t_{1,2}$ are purely imaginary.  The transition  probability $P^{++}$ can be estimated with the standard Dykhne formula as
\be
P^{++}\large =e^{-(2/b) {\rm Im} \left[ \int_{0}^{t_1} \Delta E(t) \, dt \right]}, \quad r<1,
\label{dykhne1-ppp1}
\ee
where the final integration point $t_1$ is the imaginary root in (\ref{t12}) that is closer to the real time axis.  

{\bf Phase II}: $r>1$, i.e., $e_1 e_2> g^2$.
Here, 
the branching points have both real and imaginary parts. Moreover, the imaginary parts are equal to each other, so both the branch cuts are relevant because they correspond to semiclassical evolution trajectories with comparable amplitudes. 

The transition probability is given by a generalized Dykhne formula that sums the amplitudes of both trajectories and only then takes its absolute value squared:
\be
P^{++}=\Big \vert e^{-\frac{i}{b}  \int_{0}^{t_1} \Delta E(t) \, dt +i\phi_g}+e^{-\frac{i}{b} \int_{0}^{t_2} \Delta E(t) \, dt } \Big \vert^2,
\label{dykhne1-ppp2}
\ee
where $\phi_g$ is a geometric phase difference between the two trajectories. It is  of subdominant order $O(1)$ in comparison to the integrals in (\ref{dykhne1-ppp2}), and for real Hamiltonians can take only discrete values $0$ or $\pi$. In appendix~\ref{asec-pi}, we calculate this phase for both $H_1$ and $H_2$ and show that for the symmetric model $\phi_g=0$.

The integrals in (\ref{dykhne1-ppp1}),~(\ref{dykhne1-ppp2}) cannot be simplified anymore. 
In this form, the probability $P^{++}$ is already much easier to calculate numerically than by solving the Schr\"odinger equation numerically directly. In appendix~\ref{asec-sym}, we show that the integrals can be  additionally simplified for some choices of the parameters. Such cases are useful for developing the intuition about the magnitude of $P^{++}$ and its dependence on the parameters.

The result (\ref{dykhne1-ppp2}) is valid  for $g^2/b \gg 1$. Naturally, at $g=0$ it makes an unphysical prediction: $P^{++}=2$. In order to adjust Eq.~(\ref{dykhne1-ppp2}), we  note that in our case 
$$
t_1=-t_2^*,
$$
 so, we have
$$
 {\rm Re}\left[\frac{1}{b}\int_{0}^{t_1} \Delta E(t) \, dt \right]=-{\rm Re}\left[\frac{1}{b}\int_{0}^{t_2} \Delta E(t) \, dt \right],
$$
$$
 {\rm Im}\left[\frac{1}{b}\int_{0}^{t_1} \Delta E(t) \, dt \right]={\rm Im}\left[\frac{1}{b}\int_{0}^{t_2} \Delta E(t) \, dt \right],
$$
and the desired approximation that makes $P^{++}=1$ at $g=0$ is
\be
P^{++}\approx\frac{\cos^2\left[{\rm Re}\left(\frac{1}{b}\int_{t_0}^{t_1} \Delta E(t) \, dt \right)\right]}{\cosh^2 \left[{\rm Im}\left(\frac{1}{b}\int_{t_0}^{t_1} \Delta E(t) \, dt \right) \right]}.
\label{dykhne1-ppp3-1}
\ee
Finally, using (\ref{interm-form1}), we obtain the desired approximation for the four-state model $H_1$:
\be
P_{3\rar 4} \approx \left(1-e^{-4\pi g^2/b}\right)^2\frac{\cos^2\left[{\rm Re}\left(\frac{1}{b}\int_{t_0}^{t_1} \Delta E(t) \, dt \right)\right]}{\cosh^2 \left[{\rm Im}\left(\frac{1}{b}\int_{t_0}^{t_1} \Delta E(t) \, dt \right) \right]}.
\label{psym-fin}
\ee

\begin{figure}[t!]
    \centering
    \includegraphics[width=0.45\textwidth]{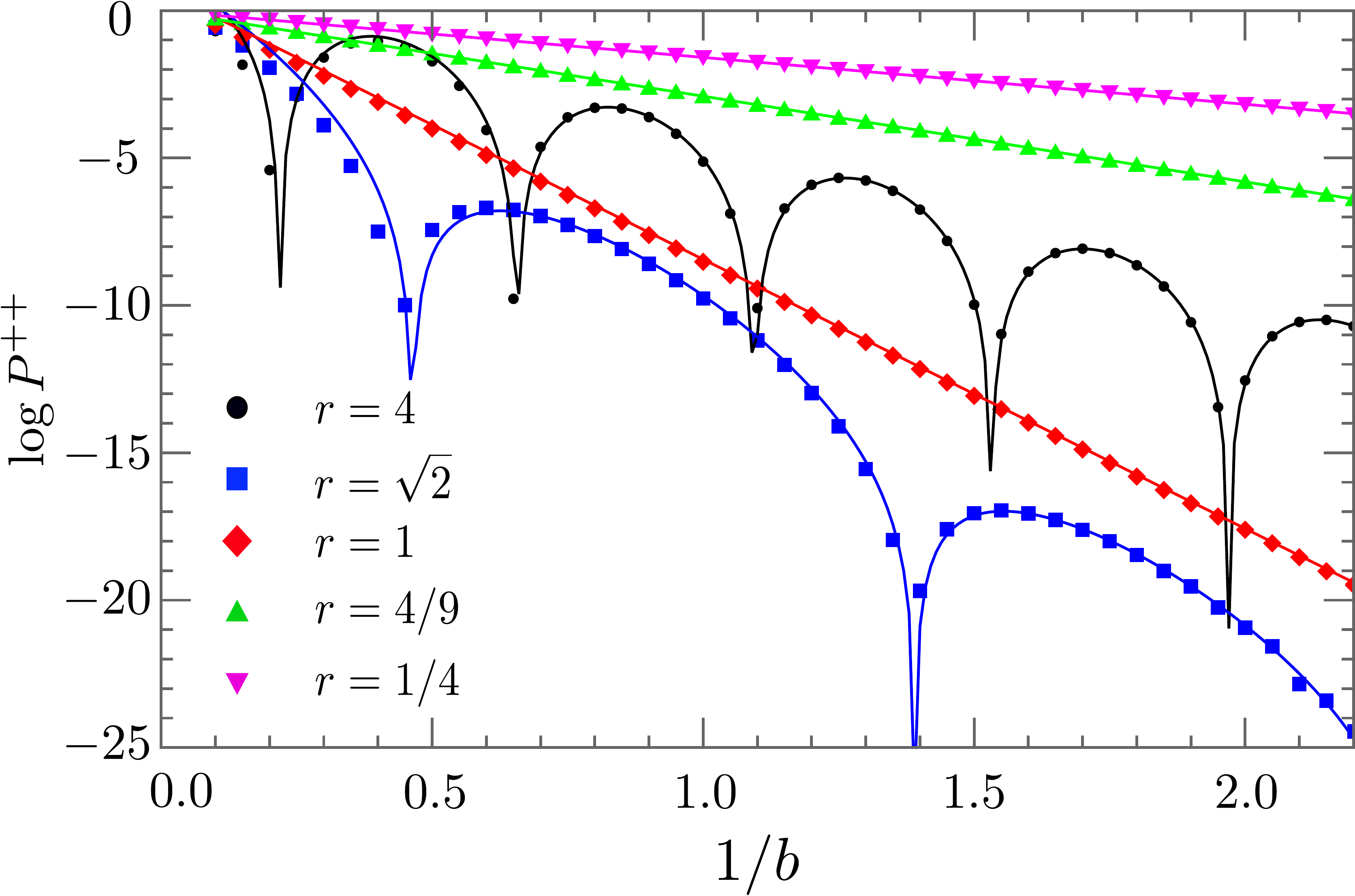}
    \caption{Probability $P^{++}$ calculated for  the symmetric model  numerically (plot markers),  and the corresponding semiclassical  prediction by Eqs.~(\ref{dykhne1-ppp1}) and (\ref{dykhne1-ppp2}) for $r=4$ (black), $r=\sqrt{2}$ (blue), $r=1$ (red), $r=4/9$ (green), and $r=1/4$ (magenta).
      }
      \label{check-fig1}
\end{figure}

In Fig.~\ref{check-fig1},  the analytical predictions  (\ref{dykhne1-ppp1}) and (\ref{dykhne1-ppp3-1}) are compared to the results  obtained by solving the Schr{\" o}dinger equation (\ref{eq-eff1})  numerically for several values of the parameter combination $e_1 e_2/g^2$. The analytical result for the critical case, $e_1e_2=g^2$ (the red curve), was taken from appendix Eq.~(\ref{e1e2g2}). 
 Phases I and II are clearly distinguishable in numerical simulations:   in  phase I,   $P^{++}$ decays monotonously with increasing  $1/b$, whereas in  
phase II, $P^{++}$  oscillates as a function of $1/b$.  

Note also that  $P^{++}$ increases with increasing coupling strength $g$ in phase I. For large $g$, the leading contribution to the sum, $P_{3\rightarrow 3}+P_{3\rightarrow 4}$, is dominated by the transition probability $P_{3\rightarrow4}$, and consequently $P^{++}$. Hence, our result agrees with the asymptotic behavior that was found for the symmetric model $H_1$ in \onlinecite{malla-LZ} in the limit $e_1e_2\ll g^2$. 


\subsection{Antisymmetric model with the Hamiltonian $H_2$}
For the Hamiltonian (\ref{ham-2}), there are  two  couplings, $g$ and $\gamma$. The asymmetry between $g$ and $\gamma$ requires from us to introduce new linear combinations of the states $|1\ra$ and $|2\ra$:
\begin{figure}[t!]
    \centering
    \includegraphics[width=0.45\textwidth]{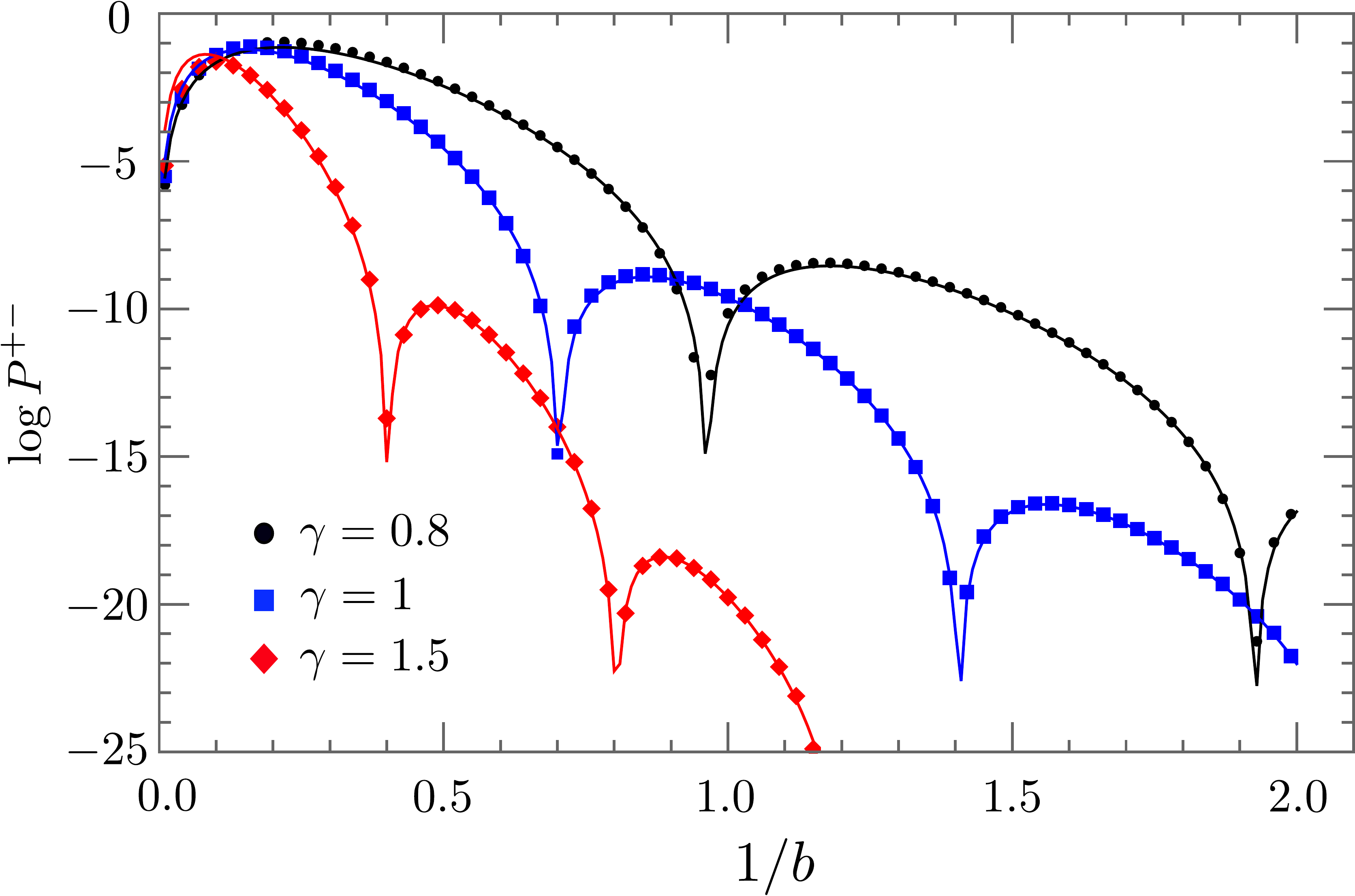}
    \caption{Probability $P^{+-}$ for the antisymmetric model, calculated numerically (plot marker) for the evolution with the Hamiltonian (\ref{eq-eff2}). Solid curves are the corresponding  semiclassical predictions by Eq.~(\ref{dykhne1-ppp3}) at $e_1=e_2=2$, $g=2$, and  $\gamma=0.8$ (black), $\gamma=1$ (blue),  $\gamma=1.5$ (red).
    \label{antisym-check1}
      }
\end{figure}
\begin{equation}
\label{newstates2}
\nonumber \vert A^+\ra=\frac{g\vert 1\ra +\gamma \vert2\ra}{\sqrt{g^2+\gamma^2}}, \quad
\vert A^-\ra=\frac{-\gamma \vert 1 \ra + g \vert 2\ra}{\sqrt{g^2+\gamma^2}},
\ee
such that $|3\ra$ couples directly to $|A^+\ra$ 
and $|4\ra$ couples to $|A^-\ra$.
In the limit $\tau \rar \infty$, both level 3 and level 4 cross the diabatic energies of  $|1\ra$ and $|2\ra$  at time moments, respectively, $t_{-}$ and $t_{+}$. 
Hence, the transition probability $P_{3\rightarrow 4}$ is
\be
P_{3\rightarrow 4}=P_{3\rightarrow A^+}P^{+-}P_{A^-\rightarrow 4},
\label{p34-2}
\ee
where $P^{+-}$ is the probability of the transition from  $|A^+\ra$ as $t\rar t_{-}$ to $|A^-\ra$ as $t\rar t_+$, and the fast nonadiabatic transitions are given by the standard LZ formula:
$$
P_{3\rightarrow A^{+}}=P_{A^-\rightarrow4}=1-e^{-2\pi(g^2+\gamma^2)/b}.
$$
In order to find  $P^{+-}$, we follow the previous approach: we rescale time and obtain the analogous to  (\ref{eq-eff1}) effective Schr\"odinger equation for evolution in the subspace  $\vert A^{\pm}\ra$ during time  interval $t\in (-1,1)$ and the effective  2$\times $2 Hamiltonian 
\be
H^{A}_{\rm eff}(t)=e_1e_2 \left( 
\begin{pmatrix}
\frac{r_{-}}{r_{+}}-\frac{r_+}{t+1} & -\sqrt{1-r_{-}^2/r_+^2}\\-\sqrt{1-r_{-}^2/r_+^2} &
-\frac{r_{-}}{r_{+}}-\frac{r_+}{t-1}
\end{pmatrix}
\right),
\label{eq-eff2}
\ee
where 
$$
r_{\pm}\equiv (g^2\pm \gamma^2)/(e_1e_2).
$$
The corresponding distance between the adiabatic levels of $H_{\rm eff}^A$ is
\be
\Delta E_A
=\frac{2e_1e_2 }{(1-t^2)} \sqrt{(t^2-1)^2+2 (t^2-1)r_{-}+r_+^2}.
\label{energy-asy1}
\ee

This expression has almost the same structure as Eq.~(\ref{en-dif1}) for the symmetric model.
Similarly, the transition from $\vert A^+\ra$ to $\vert A^-\ra$ requires a passage through the avoided crossing at $t=0$, and there are two branching points $t_1$ and $t_2$ that are   obtained by setting $\Delta E_A=0$. 

However, we found two qualitative differences of  the antisymmetric model from the symmetric one. First, the antisymmetric model does not have phase I, i.e., it always leads to the phase with two equally important branch cuts at time points
\be
t_{1,2}=\left(1-r_{-}\pm i\sqrt{r_+^2-r_{-}^2} \right)^{1/2},
\label{t12-2}
\ee
where the convention for  ``$(\ldots)^{1/2}$" is to keep only the square roots in the upper complex plane. Such roots in (\ref{t12-2}) satisfy the relation $t_1=-t_2^*$ as in phase II of the symmetric model. Hence, the antisymmetric model is always in the phase with the oscillatory behavior.
Second, we show in appendix~\ref{asec-pi} that the geometric phase for the antisymmetric model is $\phi_g=\pi$.  Hence, 
\be
P^{+-}\approx\Big \vert e^{-\frac{i}{b}  \int_{0}^{t_1} \Delta E_A(t) \, dt }-e^{-\frac{i}{b} \int_{0}^{t_2} \Delta E_A(t) \, dt } \Big \vert^2.
\label{dykhne1-ppp3}
\ee
In Fig.~\ref{antisym-check1}, we provide the  numerical check of (\ref{dykhne1-ppp3}), which confirms the analytical prediction, including the effect of the topological phase $\phi_g=\pi$.

At $g=0$, this formula predicts correctly that $P^{+-}=0$, and thus does not need further adjustments. However, we found that a slightly better fit to numerical simulations in the strongly nonadiabatic regime for the Hamiltonian $H_2$ is given by
\be
P_{3\rar 4}\approx 2\left(1-e^{-2\pi(g^2+\gamma^2)/b}\right)^2 \frac{\sin^2 \left[\frac{1}{b}{\rm Re} \int_{0}^{t_1} \Delta E_A(t) \, dt \right]}{\cosh\left[\frac{2}{b}{\rm Im}  \int_{0}^{t_1} \Delta E_A(t) \, dt \right]}.
\label{pfin}
\ee

Formulas~(\ref{psym-fin}) and (\ref{pfin}) become asymptotically exact in the adiabatic limit $1/b\rar \infty$. It is instructive to look also at how they perform in the strongly nonadiabatic regime, for which they cannot be justified rigorously.   In Fig.~\ref{4state-fig}, we show that,  although small deviations from the numerically exact predictions are generally visible, in a broad range of the parameters both formulas perform quite well. They still correctly predict multiple oscillations, including phases and amplitudes, even in the nonadiabatic regime. Thus our semiclassical theory is sufficiently rigorous for, e.g., planning the future experiments without resorting to exhaustive numerical simulations.

\begin{figure}[t!]
    \centering
    \includegraphics[width=0.45\textwidth]{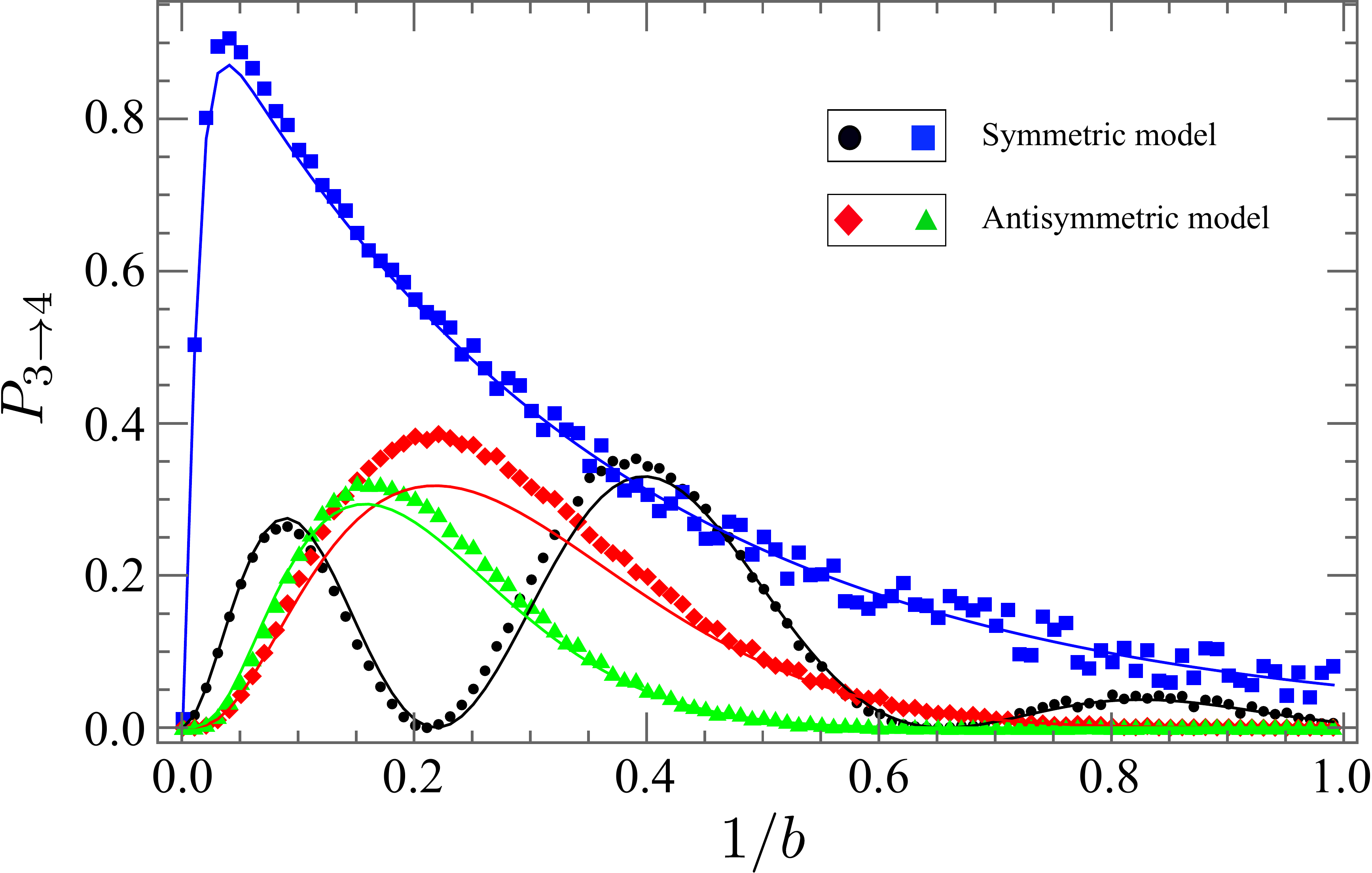}
    \caption{Transition probabilities $P_{3\rightarrow 4}$ are shown for the
    numerically exact simulations (plot markers)  and campared to the analytical approximations (solid lines) from Eqs.~(\ref{psym-fin})   and (\ref{pfin}) for both the symmetric, $H_1$, and antisymmetric, $H_2$, models.  In all cases:  $e_1=e_2=2$. The coupling in the symmetric model is $g=1$ (black) and $g=3$ (blue), and in the antisymmetric model:  $g=2$, and $\gamma=0.8$ (red) and $\gamma=1$ (green). 
      }
      \label{4state-fig}
\end{figure}

\section{Discussion}
The concept of integrability in MLZ theory originally was introduced to unify various fully solvable models. Interestingly, there are also large classes of systems that are formally integrable but cannot be fully solved. 
Thus, we showed that all LZ-grid models satisfy the integrability conditions for time-dependent Hamiltonians \cite{commute}. 
As a result, the state-to-state transition probabilities in all such models are invariant  of  parameter rescaling (\ref{invar2}) but this is not sufficient to determine these probabilities analytically.

Nevertheless, this symmetry is nontrivial and different from the set of the previously known exact results that were derived  for LZ-grids earlier \cite{usuki}. In this sense, the  integrability \cite{commute} is akin in effect to a conservation law for a time-independent Hamiltonian. 

We are not aware of experimental studies of such  symmetries in explicitly time-dependent Hamiltonians, although these symmetries should be realizable in a broad class of physical systems \cite{li-dynamic}. Given that the Landau-St\"uckelberg interferometry for a four-state antisymmetric model (\ref{ham-2}) has been already demonstrated in quantum dots \cite{petta-4lz1,petta-4lz2}, the  integrability of LZ-grids can also  be  tested. Specifically, for the Hamiltonian (\ref{ham-2}), we predict nontrivial but simple-looking relations between the state-to-state transition probabilities  (\ref{rel1})-(\ref{rel3}) in the dynamics induced  by a linear potential chirp.

The experiment to observe such a nontrivial symmetry requires a modification of the control and measurement protocol in \cite{petta-4lz1}. The standard Landau-St\"uckelberg interferometry protocol induces periodic transitions through the region with nonadiabatic transitions.  It leads to spectacular multi-dimensional plots for the state probability dependence on the potential sweeping rate and the duration of the periodic control. However, such measurements are influenced by all parameters of the scattering matrix for a single passage through the nonadiabatic region.  Hence, an analytical interpretation of such patterns for multi-state quantum systems is usually problematic due to the large number of relevant parameters. 

Here, we propose to explore the multistate Landau-Zener transition probabilities  directly, for example,  resetting the state and then measuring it  after each linear potential  chirp in a coupled quantum dot system \cite{petta-4lz1}. The advantage of such an experiment would be
the direct measurement of the characteristics, for which either exact or semiclassical analytical predictions can be established. By comparing the experimental data to the theory, one can then explore such phenomena as the probability oscillations due to completely nonadiabatic interference effects that we exposed, as well as to confirm the exact constraints that follow from the model's integrablity. 

On the side of mathematical physics, we found further confirmations to the conjecture in \cite{quadr-LZ} that the time-dependent Hamiltonian integrability often leads to a semiclassical approach to calculate the leading contributions to the overgap transition amplitudes in multistate systems. 
This approach is an opportunity to explore the driven quantum models with considerable complexity. Although generally approximate, it produces reasonable approximations even in the strongly nonadiabatic regime, in which it cannot be rigorously justified. 

Our semiclassical analysis reveals that even the simplest driven models can show complex behavior, such as phase transitions, in which the adiabatic limit plays a similar role to the thermodynamic limit in many-body systems.  Thus, in the symmetric model, the interfering semiclassical trajectories do not necessarily lead to oscillations of measurable characteristics. Rather there are two phases: one with oscillatory behavior and another one with a monotoneous decay of the transition probability. We also found that models with similar structure and similar oscillatory behavior, can belong to different topological classes that are characterized by the topological phase $\phi_g$ between the interfering semiclassical trajectories. The integrability allows us to observe and quantify such phenomena in relatively complex and strongly driven quantum systems.

\appendix
\section{Topological $\pi$-phase}
\label{asec-pi}

The $\pi$-phase between the amplitudes of different semiclassical trajectories in complex time has been known in  literature, see, e.g., in \cite{joye-prefactor}. In practice, however, this phase has been usually inferred numerically, and we do not know of a published rigorous analytical theory for its calculation. Therefore, here we provide such a theory, which exposes the topological nature of the $\pi$-phase and provides a simple path for its calculation. Our models with the Hamiltonians $H_{\rm eff}^s$ and $H_{\rm eff}^A$ are particularly suitable for illustration because all calculations for them can be performed analytically, and the results differ for the different models.

The geometric phase appears generally in quantum mechanics when evolution is considered along a closed path in the parameter space.
In the context of the Dykhne formula, such a path can be found
if we note that the phase difference between the amplitudes of the trajectories, ${\bm C}^{-}$ and ${\bm C}^+$ in Fig.~\ref{paths-fig}(a), that go through two different branching points, is the same as the phase acquired during the evolution along a path ${\bm C}=\left({\bm C}^{-}\right)^{-1} {\bm C}^{+}$, that is, the cyclic trajectory on the Riemann surface that starts and ends at $t=0$ on the original real time axis. ${\bm C}$  follows ${\bm C}^{+}$ along its direction marked in Fig.~\ref{paths-fig}(a), and then switches to ${\bm C}^-$ but follows it in the opposite direction to what is marked in Fig.~\ref{paths-fig}(a). In Fig.~\ref{paths-fig}(b) we show that this path is a knot that winds around the origin points of the branch cuts. 
\begin{figure}[t!]
    \centering
    \includegraphics[width=0.45\textwidth]{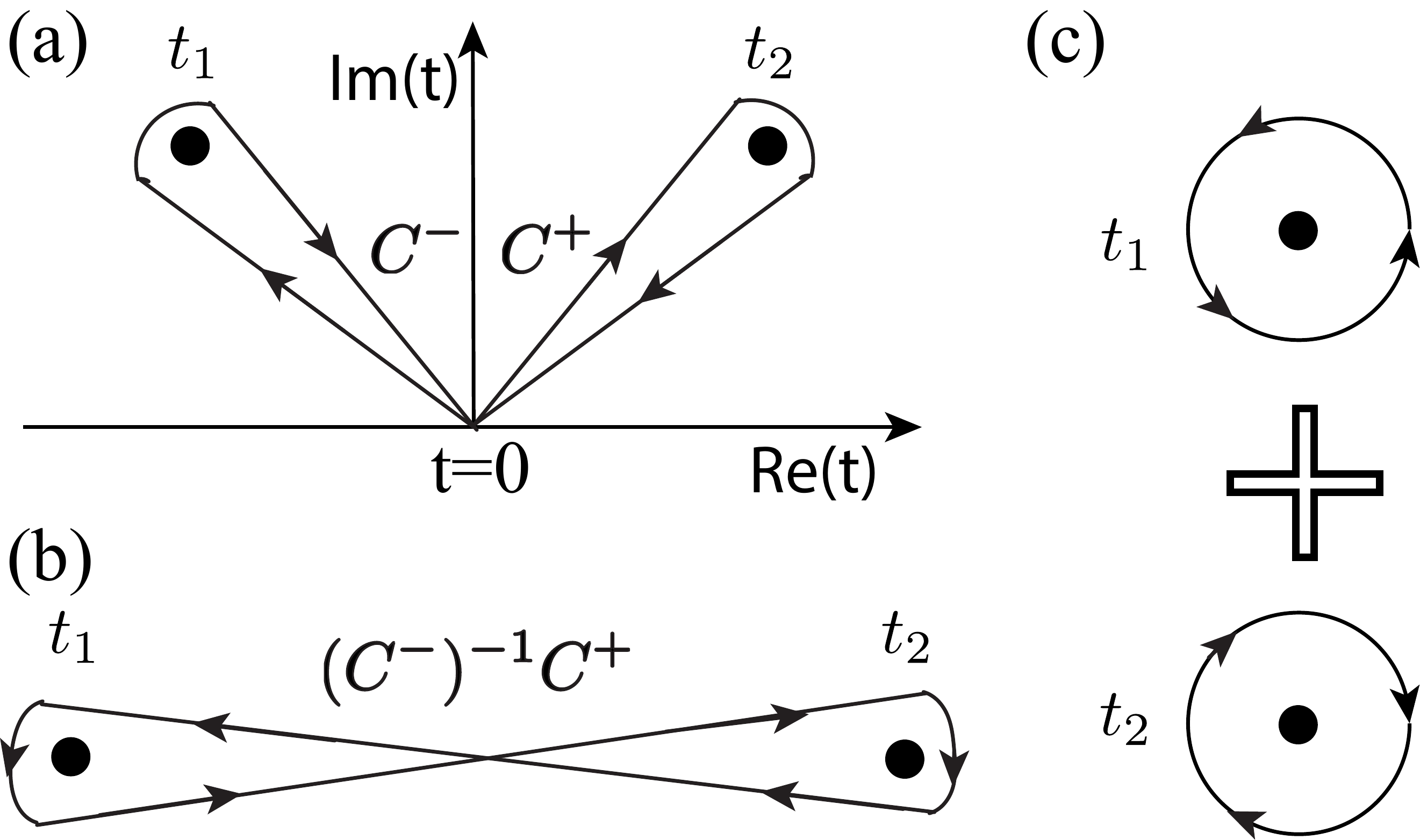}
    \caption{ (a) The two trajectories, ${\bm C}^{-}$ and ${\bm C}^{+}$, around the corresponding branch points $t_1$ and $t_2$, that make comparable contributions  to the semiclassical probability amplitude. (b) The path ${\bm C}=({\bm C}^{-})^{-1}{\bm C}^{+}$ is topologically equivalent to a knot that winds around the branching points $t_{1,2}$. (c) The integral over the knot path in Eq.~(\ref{toppase1}) is the sum of contributions from the infinitely small circular contours that wind in opposite directions around the poles at $t_{1,2}$.}
      \label{paths-fig}
\end{figure}

Any real two-state time-dependent Hamiltonian can be written in the form 
\be
H(t)=f(t){1}_2 + Z(t) \sigma_z +X(t) \sigma_x,
\label{twoH-3}
\ee
where $f(t)$, $Z(t)$, and $X(t)$ are functions of time. Let us then define 
\be
\sin \theta =\frac{X(t)}{\sqrt{X(t)^2+Z(t)^2}},\quad \cos \theta =\frac{Z(t)}{\sqrt{X(t)^2+Z(t)^2}}. \quad 
\label{coss}
\ee

Along the cyclic path in complex plane, ${\bm C}$, an eigenstate of $H(t)$ can still be parametrized by an angle $\theta$:
\be
|u(\theta) \ra = \left( \begin{array}{c}
-\sin \theta/2 \\
\cos \theta/2
\end{array} \right).
\label{realH-D1}
\ee
The topological phase originates from the fact that the Hamiltonian is periodic function of $\theta$, whereas the eigenstate vector (\ref{realH-D1}) depends on $\cos (\theta/2)$ and $\sin (\theta/2)$. Hence, if  $\theta$ changes from $0$ to $2\pi$ during an adiabatic evolution,  the Hamiltonian returns to its initial form at the end but the state vector can change  sign:
\be
|u(2\pi)\ra_{\theta(\bm C)=2\pi} = - |u (0)\ra.
\label{sign-ch-D1}
\ee
In other words, the state vector acquires a phase $\pi$.  Generally, a periodic adiabatic evolution of  parameters  changes the angle $\theta$  by $2\pi n$, where $n$ is an arbitrary integer. 
The topological phase is then either $0$, if $n$ is even, or $\pi$, if $n$ is odd. There cannot be another type of a geometric phase in this case because  $\la u(\theta)|\frac{\partial u(\theta)}{\partial \theta} \ra=0$. 


For a closed trajectory ${\bm C}$, which can be parametrized by time $t$, the topological phase is given by 
\be
\phi_g=\frac{1}{2}\int_{\bm C} \frac{d\theta}{d t} dt. 
\label{toppase1}
\ee
Note that 
$$
\frac{d\theta}{dt}= \frac{1}{\cos \theta} \frac{d \sin \theta}{dt}.
$$
For the symmetric model (\ref{eq-eff1}), we have 
$$
Z=\frac{2g^2t}{(t^2-1)}, \quad X=e_1e_2,
$$
so, 
\be
\frac{1}{2}\frac{d\theta}{dt}=\frac{r(t^2+1)}{4t^2+r^2(1-t^2)^2} = \frac{(t^2+1)}{r(t^2-t_1^2)(t^2-t_2^2)},
\label{dt1}
\ee
where $r=e_1e_2/g^2$, and $t_{1,2}$ are the two roots in the upper complex half-plane, which are written in  (\ref{t12}).

Note that the integrand in (\ref{dt1}) has simple poles at $t_{1,2}$ rather than the branching points. Hence, the integral over ${\bm C}$ is given by the  difference of the residues at these poles, as illustrated in Fig.~\ref{paths-fig}(c):
\be
\phi_g=\pi i \left[ Res\left(\frac{d\theta}{dt}\right)_{t_1}- Res\left(\frac{d\theta}{dt}\right)_{t_2} \right],
\label{ph-res}
\ee
where $Res(\ldots)_a$ is the residue of the expression at a simple pole $a$, and  where the minus sign is because  ${\bm C}$ winds around $t_1$ and $t_2$ in opposite directions. Substituting the roots from (\ref{t12}) to (\ref{ph-res}), and making sure that $t_1=-t_2^*$, we find for the effective two-state system with the Hamiltonian $H_{\rm eff}^s$  that 
\be
\phi_g=0.
\label{phisym}
\ee

Analogously, for the effective Hamiltonian (\ref{eq-eff2}) that corresponds to the antisymmetric model (\ref{ham-2}), we have 
\be
\frac{d\theta}{dt}=\frac{-2\sqrt{r_+^2-r_{-}^2} t}{(t^2-t_1^2)(t^2-t_2^2)} ,
\label{dt2}
\ee
where $t_{1,2}$ are  given by (\ref{t12-2}), and $t_{1}=-t_2^*$. Substituting (\ref{dt2}) and (\ref{t12-2}) into (\ref{ph-res}), we find that for the effective Hamiltonian  (\ref{eq-eff2}) the topological phase is 
\be
\phi_g=\pi.
\label{phi-ants}
\ee
Thus, despite  similar oscillatory behavior, the symmetric and antisymmetric models are characterized by the different values of  topological phase $\phi_g$. In this sense the models belong to different topological classes.

\section{Analytically simple special cases}
\label{asec-sym}
\subsection{Symmetric model: the critical point at $e_1e_2=g^2$}
The condition $e_1 e_2=g^2$ marks the phase transition point between phase I and phase II. In this special case the two  branching points are degenerate and purely imaginary,  i.e., $t_1=t_2=i$.  The  adiabatic energy difference simplifies to 
$$
\Delta E = 2g^2\frac{t^2+1}{1-t^2}.
$$
Substituting $t=ix$ into the integral in the Dykhne formula (\ref{dykhne1-ppp1}) we find 
$$
2{\rm Im} \left[ \int_0^{t_1} \Delta E \, dt \right] = 2g^2 \int_0^1 \frac{1-x^2}{1+x^2} \, dx = 2(\pi -2).
$$
Following \cite{joye-prefactor}, we should set the exponential prefactor in this special case to be 2 rather than 1 because  the behavior near the zero is  $\Delta E\sim (t-t_1)$ rather than the typical $\Delta E \sim (t-t_1)^{1/2}$.  This leads to the transition probability
\be
P^{++}_{e_1e_2=g^2} = 2e^{-2(\pi-2)g^2/b}.
\label{spcl-p++}
\ee
This formula is valid only asymptotically in the limit $g^2/b \gg 1 $. Without affecting the behavior in  this domain, we can also write
\be
P^{++}_{e_1e_2=g^2} \approx \frac{1}{\cosh (2(\pi-2)g^2/b)},
\label{e1e2g2}
\ee
which is still generally approximate but has right value, $1$, at $g=0$. Finally, the semiclassical formula for the transition probability in the original 4-state model for this special case is
\be
P_{3\rar 4}^{e_1e_2=g^2} \approx \frac{ \left(1 -e^{-4\pi g^2/b} \right)^2}{\cosh (2(\pi-2)g^2/b)}.
\label{equalgP11}
\ee

\subsection{Symmetric model: the case with oscillations at $e_1 e_2=g^2\sqrt{2}$}
Substituting
$
e_1e_2=g^2\sqrt{2}
$, 
 to Eq.~(\ref{en-dif1}), we find
$$
\Delta E = 2\sqrt{2}g^2 \frac{\sqrt{t^4+1}}{1-t^2}.
$$
The two branch points in the upper plane are then given by
$$
t_{1} = e^{i3\pi/4}, \quad t_2=e^{i\pi/4}.
$$

To calculate the integral, we set
$$
t=e^{i\pi/4}s,
$$
 which leads to: 
$$
\frac{1}{b}\int_0^{t_2} dt \Delta E = \frac{2\sqrt{2}e^{i\pi/4}g^2}{b}\int_0^1 ds \, \frac{\sqrt{1-s^4}}{1-is^2}=\frac{g^2(
\xi+i\zeta)}{b},
$$
where the analytic expressions for $\xi$ and $\zeta$ can be written in terms of the known special functions but the expressions are a bit lengthy. We just give their numerical values, which can be found to very high precision:
$$
\xi = 1.19814\ldots, \quad \zeta = 1.9434\ldots. 
$$
The analogous calculation leads to 
$$
\frac{1}{b}\int_0^{t_1} dt \Delta E = \frac{g^2(-\xi+i\zeta)}{b},
$$
and then to
$$
P^{++}\approx \frac{\cos[\xi g^2/b]^2}{\cosh[\zeta g^2/b]^2},
$$
which provides the leading exponent for small $b$ and is $1$ at $g=0$.

\subsection{Antisymmetric model: the case with $e_1 e_2=g^2-\gamma^2$}

In the antisymmetric model there is also a special case
$$
e_1 e_2=g^2-\gamma^2, \quad r_{-}=1,
$$
such that the integrals can be computed explicitly. Substituting  $r_{-}=1$ in (\ref{energy-asy1}), we find the eigenvalue difference
$$
\Delta E_A=\frac{2e_1e_2}{(1-t^2)}\sqrt{t^4-1+r_{+}^2},
$$
and the branch points in the upper half plane are given by
$$
t_1=|r_{+}^2-1|^{1/4}e^{i3\pi/4}, \quad t_2=|r_{+}^2-1|^{1/4}e^{i\pi/4}.
$$

In order to calculate the integral we set
$$
t=|r_{+}^2-1|^{1/4}e^{i\pi/4}s,
$$
and find
\begin{widetext}
$$
\frac{1}{b} \int_0^{t_2} dt \Delta E_A = \frac{2e_1e_2 |r_{+}^2-1|^{3/4} e^{i\pi/4}}{b}\int_0^1 ds \, \frac{\sqrt{1-s^4}}{1-i|r_{+}^2-1|^{1/2}s^2}
=\frac{e_1e_2}{b}\{F[|1-r_{+}^2|]+i G[|1-r_{+}^2|]\},
$$
\end{widetext}
where analytical expressions for $F[x]$ and $G[x]$ can be written in terms of the known special functions, which can be calculated with high precision. Similarly, we evaluate the integral for the branching point $t_1$:
$$
\frac{1}{b}\int_0^{t_1} dt \Delta E_A = \frac{e_1e_2}{b}\{-F[|1-r_{+}^2|]+i G[|1-r_{+}^2|]\}.
$$
Combining the two contributions, and taking into account the topological phase $\phi_g=\pi$, we finally find:
$$
P^{+-}\approx \frac{\sin[e_1e_2F[|1-r_{+}^2|] /b]^2}{\cosh[e_1e_2G[|1-r_{+}^2|] /b]^2}.
$$

\section*{Acknowledgements}
This work was supported by the U.S. Department of Energy, Office of Science, Basic Energy
Sciences, Materials Sciences and Engineering Division, Condensed Matter Theory Program. R.K.M. was supported by LANL LDRD program under project number 20190574ECR and Center for Nonlinear Studies under project number 20190495CR.


\begin{thebibliography}{100}


\bibitem{be} S. Brundobler, and V. Elser. \textit{``S-matrix for generalized Landau-Zener problem"}, J. Phys. A {\bf 26}, 1211 (1993).


\bibitem{harmin1} D. A. Harmin and P. N. Price. {\it
Incoherent time evolution on a grid of Landau-Zener anticrossings.} Phys. Rev. A {
\bf 49}, 1933 (1994).
\bibitem{harmin2} D. A. Harmin. {\it Coherent time evolution on a grid of Landau-Zener anticrossings.} Phys. Rev. A 
{\bf 56}, 232  (1997).

\bibitem{optics-lzgrid}
D. Bouwmeester, I. Marzoli, G. P. Karman, W. Schleich, and J. P. Woerdman.
{\it Optical Galton board}. Phys. Rev. A {\bf 61}, 013410 (1999).

\bibitem{usuki} T. Usuki. {\it Theoretical study of Landau-Zener tunneling at the M+N level crossing.} Phys. Rev. B {\bf 56}, 13360 (1997).

\bibitem{sinitsyn-bath} N. A. Sinitsyn and N. Prokof'ev. 
{\it Nuclear spin bath effects on Landau-Zener transitions in nanomagnets.} Phys. Rev. B {\bf 67}, 134403 (2003)

\bibitem{hanggie} K. Saito, M. Wubs, S. Kohler, Y. Kayanuma, and P. Hanggi. {
\it Dissipative Landau-Zener transitions of a qubit: bath-specific and universal behavior}. Phys. Rev. B {
\bf 75}, 214308 (2007).

\bibitem{ashhab-lz} S. Ashhab. {\it Landau-Zener transitions in a two-level system coupled to a finite-temperature harmonic oscillator}.
Phys. Rev. A {\bf 90}, 062120 (2014).


\bibitem{garanin-grid} D. A. Garanin, R. Neb, and R. Schilling. {\it
Effect of environmental spins on Landau-Zener transitions.} Phys. Rev. B {\bf 78}, 094405 (2008).

\bibitem{recent-grid1} M. Werther,   F. Grossmann, Z. Huang, and   Y Zhao. {\it Davydov-Ansatz for Landau-Zener-Stueckelberg-Majorana transitions in an environment: Tuning the survival probability via number state excitation.} J. Chem. Phys. {\bf 150}, 234109 (2019).

\bibitem{recent-grid2} F. Bello, N. Kongsuwan, J. F. Donegan, and O. Hess. {\it Controlled Cavity-Free, Single-Photon Emission and Bipartite Entanglement of Near-Field-Excited Quantum Emitters}. Nano Lett.  {\bf 20}, 8, 5830 (2020).



\bibitem{osc-grid1} M.Kervinen, J. E. Ramírez-Munoz, A. Valimaa, and M. A. Sillanpaa. {\it Landau-Zener-Stückelberg interference in a multimode electromechanical system in the quantum regime}. Phys. Rev. Lett. {\bf 123}, 240401 (2019).

\bibitem{osc-grid2} Z. Sun, J. Ma, X. Wang, and F. Nori. {\it  Photon-assisted Landau-Zener transition: Role of coherent superposition states.} Phys. Rev. A {\bf 86}, 012107 (2012).

\bibitem{osc-grid3} P. Y. Wen, {\it et al.}. {\it Landau-Zener-Stückelberg-Majorana interferometry of a superconducting qubit in front of a mirror}. Phys. Rev. B {\bf 102}, 075448 (2020).

\bibitem{osc-grid4} R. K. Malla and M. E. Raikh. {\it Landau-Zener transition in a two-level system coupled to a single highly excited oscillator}. Phys. Rev. B {\bf 97}, 035428 (2018)

\bibitem{ostrovsky-grid} V. N. Ostrovsky, M. V. Volkov, J. P. Hansen, and S. Selsto. {\it Four-state nonstationary models in multistate Landau-Zener theory.}
Phys. Rev. B {\bf75}, 014441 (2007)


\bibitem{yurovsky-grid} V. A. Yurovsky and A. Ben-Reuven. {\it Curve crossing in linear potential grids: The quasidegeneracy approximation}.
Phys. Rev. A {\bf 63}, 043404 – Published 6 March 2001



\bibitem{quest-LZ} N. A. Sinitsyn, and V. Y. Chernyak. \textit{The quest for solvable multistate Landau-Zener models}. J. Phys. A: Math. Theor. {\bf 50}, 255203 (2017). 

\bibitem{commute} N. A. Sinitsyn, E. A. Yuzbashyan, V. Y. Chernyak, A. Patra, and C. Sun. {\it Integrable time-dependent quantum Hamiltonians.} Phys. Rev. Lett. {\bf 120}, 190402 (2018). 


\bibitem{quadr-LZ} V. Y. Chernyak,  and N. A. Sinitsyn. {\it Integrability in the multistate Landau-Zener model with time-quadratic commuting operators}. arXiv:2006.15144, submitted to J. Phys. A (2021).



\bibitem{laser-LZ} C. Sun, V. Y. Chernyak, A. Piryatinski, and N. A. Sinitsyn. {\it Cooperative Light Emission in the Presence of Strong Inhomogeneous Broadening}. Phys. Rev. Lett. {\bf 123}, 123605 (2019).
\bibitem{bcs} F. Li, V. Y. Chernyak, and N. A. Sinitsyn. {\it Quantum annealing and thermalization: insights from integrability}. Phys. Rev. Lett. {\bf 121}, 190601 (2018). 
\bibitem{gamma-LZ} V. Y. Chernyak, N. A. Sinitsyn, and C. Sun. {\it Dynamic spin localization and 
gamma-magnets.} Phys. Rev. B {\bf 100}, 224304 (2019). 


\bibitem{yuzbashyan-LZ} E. A. Yuzbashyan.  {\it Integrable time-dependent Hamiltonians, solvable Landau-Zener models and Gaudin magnets. } Ann. Phys. {\bf 392}, 323 (2018). 


\bibitem{note} Although the LZ formula should be credited to four authors: Landau, Majorana, Zener, and St\"uckelberg, the interference of semiclassical trajectories,  in the context of nonadiabatic transitions, was discussed  only by Landau and St\"uckelberg. 


\bibitem{petta-4lz1} F. Ginzel, A. R. Mills, J. R. Petta, G. Burkard. {\it Spin shuttling in a silicon double quantum dot}. Phys. Rev. B {\bf 102}, 195418 (2020).
\bibitem{petta-4lz2} X. Mi, S. Kohler, and J. R. Petta. {\it Landau-Zener interferometry of valley-orbit states in Si/SiGe double quantum dots}. Phys. Rev. B {\bf 98}, 161404(R)(2018).


\bibitem{faddeev-book} L. D. Faddeev, and L. A. Takhtajan. {\it Hamiltonian Methods in the Theory of Solitons}. (Springer, 1987).

\bibitem{patra-LZ}  A. Patra, and E. Yuzbashyan. \textit{Quantum integrability in the multistate Landau-Zener problem}.  J. Phys. A {\bf 48}, 245303  (2015).

\bibitem{malla-LZ} R. K. Malla, and M. E. Raikh. {\it Loss of adiabaticity with increasing tunneling gap in nonintegrable multistate Landau-Zener models}. Phys. Rev. B {\bf 96}, 115437 (2017).

\bibitem{four-LZ} N. A. Sinitsyn. {\it Solvable four-state Landau-Zener model of two interacting qubits with path interference}. 
Phys. Rev. B {\bf 92}, 205431 (2015).


\bibitem{li-dynamic} F. Li, and N. A. Sinitsyn. {\it Dynamic Symmetries and Quantum Nonadiabatic Transitions}. Chem. Phys. {\bf 481}, 28 (2016).

\bibitem{nogo-LZ} N. A. Sinitsyn. {\it Counterintuitive transitions in the multistate Landau-Zener problem with linear level crossings}. 
J. Phys. A: Math. Gen. {\bf 37}  10691 (2004).



\bibitem{kiselev-LZ}  M. N. Kiselev, K. Kikoin, and M. B. Kenmoe. {\it SU(3) Landau-Zener interferometry}.  EPL {\bf 104} 57004 (2013).






 

\bibitem{approx-LZ5} K. Fukushima, and T. Shimazaki. {\it Lefschetz-thimble inspired analysis of the Dykhne-Davis-Pechukas method and an application for the Schwinger Mechanism.} Ann. Phys. {\bf 415}, 168111 (2020).


\bibitem{approx-LZ0} J.‐T. Hwang, and P. Pechukas. {\it The adiabatic theorem in the complex plane and the semiclassical calculation of nonadiabatic transition amplitudes.} J. Chem. Phys. {\bf 67}, 4640 (1977).




\bibitem{joye-prefactor} A. Joye. {\it Non-trivial prefactors in adiabatic transition probabilities induced by high-order complex degeneracies.} J. Phys. A.: Math. Gen. {\bf 26} 6517 (1993).















\end{thebibliography}
\end{document}